\documentclass[a4paper,11pt]{article}
\usepackage{pos}
\usepackage{physics}
\usepackage{booktabs}
\usepackage{bm}
\usepackage[export]{adjustbox}

\newcommand{\pizero}{{\pi^0}}
\newcommand{\piplus}{{\pi^+}}

\title{Electromagnetic pion mass splitting using PV-regulated photon propagator}

\author*[a,b,c]{Alessandro De Santis}
\author[d]{Dominik Erb}
\author[a,d,f]{Harvey B. Meyer}

\affiliation[a]{Helmholtz Institut Mainz, Staudingerweg 18, D-55128 Mainz, Germany}
\affiliation[b]{Johannes Gutenberg-Universit{\"a}t Mainz, 55099 Mainz, Germany}
\affiliation[c]{GSI Helmholtz Centre for Heavy Ion Research, 64291 Darmstadt, Germany}
\affiliation[d]{PRISMA$^+$ Cluster of Excellence \& Institut f{\"u}r Kernphysik, Johannes Gutenberg-Universit{\"a}t Mainz,
D-55099 Mainz, Germany}
\affiliation[f]{Theoretical Physics Department, CERN, 1211 Geneva 23, Switzerland}

\emailAdd{desantia@uni-mainz.de}

\abstract{
Several hadronic observables are nowadays computed in lattice QCD with a sub-percent precision which requires the inclusion of strong isospin-breaking and electromagnetic effects. Most of the methods that implement the photon propagator in finite-volume lead to power-law suppressed finite-size effects and do not allow for a straightforward crosscheck against phenomenology and other calculations. Both issues can be avoided by working with a Pauli-Villars regulated photon propagator defined directly in the continuum and infinite volume. This methodology can be profitably exploited to improve the determination of leading-order electromagnetic corrections to several observables such as the HVP or nucleon masses. In this work  we apply the strategy to the charged/neutral pion mass difference using CLS ensembles.  }

\FullConference{The 42nd International Symposium on Lattice Field Theory (LATTICE2025)\\
2-8 November 2025\\
Tata Institute of Fundamental Research, Mumbai, India\\}

\begin{document}
\maketitle

\section{Introduction}

The ever-increasing precision of first principles lattice QCD calculations has now reached a level at which the determination of electromagnetic effects, neglected for a long time, is essential for many hadronic observables. Calculating them entails several technical and computational challenges. One of the main issues is that the prescription to include QED effects into lattice QCD calculations is scheme and formulation dependent. As a consequence, cross-checks at intermediate stages among different groups and comparisons with phenomenological predictions are hardly possible, which represents a drawback since most of the lattice calculations proceed incrementally, one class of diagrams at a time. In addition, the insertion of a photon propagator, present in diagrams stemming from the expansion of the QCD action in power of the QED coupling $\alpha_\mathrm{em}$ \cite{deDivitiis:2013xla}, usually leads to UV divergences if the proper counterterms are not taken into account.

In \cite{Biloshytskyi:2022ets}, a computational strategy aimed at overcoming this problem was proposed in the context of the anomalous magnetic moment of the muon, for which  the determination of isospin-breaking and QED effects plays a prominent role. The formalism relies on replacing the internal photon line by a Pauli-Villars (PV) regulated propagator in infinite volume with $\Lambda$ being the photon virtuality cutoff scale.  Although this procedure introduces an additional scale $\Lambda$ into the calculation flow, which eventually requires a numerical $\Lambda \to \infty$ limit, the practical advantages are quite evident: (i) the regularized photon propagator is independent of the specific lattice QCD implementation making the comparison between groups adopting different lattice QCD actions and with phenomenological predictions more clear and unambiguous; (ii) using a photon propagator already in infinite volume avoids power-law finite-volume effects, which arise in the commonly used QED$_L$ approach (see for instance \cite{Hayakawa:2008an}).

In this work we test this method in the case of the charged/neutral pion mass difference $\Delta M_\pi=M_\piplus-M_\pizero$, the pion mass spliting in short, by performing a complete calculation at several values of the lattice spacing, pion masses and photon cutoff scales. We also show that, working with a scale $\Lambda$ allows for a direct comparison with the elastic contribution to $\Delta M_\pi$  calculated using the Cottingham formula \cite{Stamen:2022uqh}, which expresses the mass splitting in terms of the forward Compton amplitude.

\section{Formalism}
\begin{figure}
\centering
\includegraphics[width=0.48\linewidth]{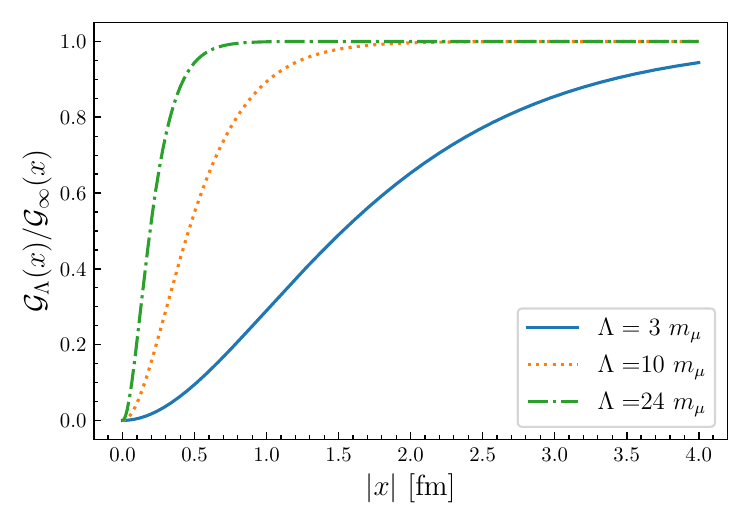}
\includegraphics[width=0.48\linewidth]{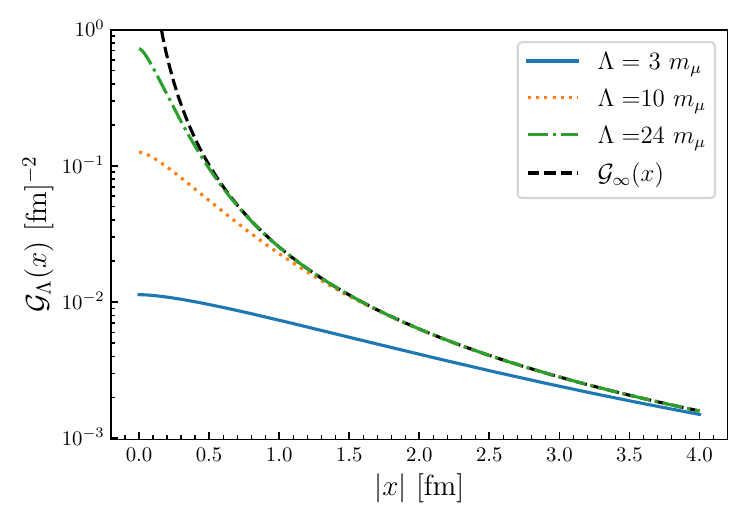}
\caption{\emph{Left plot:} ratio between the PV-regulated photon propagator and the unregulated one ($\Lambda=\infty$) for different values of the photon mass. \emph{Right plot:} comparison between the PV-regulated photon propagator and the unregulated one. }
\label{fig:photon_propagator}
\end{figure}

Mesons and baryons mass splittings can be determined by taking the ground state contribution of Euclidean correlation functions in the QCD+QED theory. The QCD part is determined stochastically via Monte Carlo simulations while the QED terms are treated perturbatively by an expansion in powers of the electromagnetic coupling \cite{deDivitiis:2013xla}. In particular, we follow the same strategy and setup that are used in \cite{Erb:2025nxk} and write the leading-order electromagnetic contribution to the pion mass splitting using the so-called \textit{mid-point} method,
\begin{flalign}\label{eq:midpoint}
\Delta M_\pi(\Lambda) =\frac{e^2}{2}\lim_{z_0\to \infty} \frac{\int \dd[4]x\; \delta^{\mu\nu}\mathcal{G}_\Lambda(x) \expval{O_\pi\big(\frac{z_0}{2}\big)J_\mu^\mathrm{em}(x)J_\nu^\mathrm{em}(0)O_\pi^\dagger\big(\frac{-z_0}{2}\big)}_\mathrm{QCD}}{\expval{O_\pi\big(\frac{z_0}{2}\big)O_\pi^\dagger\big(-\frac{z_0}{2}\big)}_\mathrm{QCD}}.
\end{flalign}
In the previous expression $e^2=4\pi\alpha_\mathrm{em}=4\pi/137.036$, $O_\pi(x)$ is the interpolating operator with quantum numbers of the pion and $J_\mu^\mathrm{em}$ is the light-quark electromagnetic vector current.  The subscript QCD stresses that these expectation values are taken with respect to only QCD gauge fields. We have then introduced the following  doubly PV-regulated photon propagator in Feynman gauge,
\begin{flalign}\label{eq:PV_propagator}
\mathcal{G}_\Lambda(x) = \frac{1}{4\pi^2 |x|} -2G_{\frac{\Lambda}{\sqrt{2}}}(x)+G_\Lambda(x),
\end{flalign}
where $G_m(x)=mK_1(m|x|)/[4\pi^2|x|]$ is the propagator of a massive scalar and $K_1(x)$ is the modified Bessel function of the second kind. The PV-propagator is defined in the infinite volume and in the coordinate space and the photon mass $\Lambda$ guarantees that $\mathcal{G}_\Lambda(x)$ is finite in the $|x|\to 0$ limit.  It is shown, together with the unregularized one ($\Lambda=\infty$), in Fig.~\ref{fig:photon_propagator}. For convenience we express the photon mass as factors of the muon mass $m_\mu=0.10566~$GeV. In order to keep cutoff effects small, we should have in general $a\Lambda\ll 1$. Nevertheless, relying upon the special fact the $\Delta M_\pi$ is naturally UV-finite, we consider in this work photon UV cutoff from $3\,m_\mu$ up to the largish value $120\,m_\mu\sim 12~$GeV.
\section{Numerical setup}
\begin{figure}
\centering
\includegraphics[width=0.48\linewidth,valign=c]{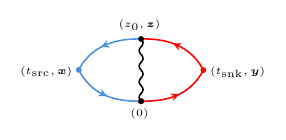}
\includegraphics[width=0.48\linewidth,valign=c]{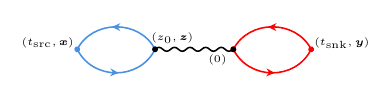}
\caption{The connected (left) and disconnected (right) diagrams relevant to the determination of $\Delta M_\pi$ at leading-order in $\alpha_\mathrm{em}$. See \cite{deDivitiis:2013xla} for the derivation.}
\label{fig:diagrams}
\end{figure}

On the lattice, the evaluation of the fermionic Wick contractions of the expectation value appearing in Eq.~\ref{eq:midpoint} leads to two diagrams with different topology, a quark connected one and a quark disconnected one, which are shown in Fig.~\ref{fig:diagrams}. Up to the insertion of the photon propagator, these consist of the following quark contractions,
\begin{flalign}
C_{\mu\nu}^\mathrm{conn}(z_0,t_\mathrm{sep},\bm{x},\bm{y},\bm{z})&=\Tr\big[\gamma_\mu S(0,x)\gamma_5 S(x,z)\gamma_\nu S(z,y)\gamma_5 S(y,0)\big], \\
C_{\mu\nu}^\mathrm{disc}(z_0,t_\mathrm{sep},\bm{x},\bm{y},\bm{z})&=\Tr\big[\gamma_\mu S(0,y)\gamma_5 S(y,0)]\times \Tr\big[\gamma_\nu S(z,x)\gamma_5 S(x,z)\big], 
\end{flalign}
where $S(u,v)$ is the fermion propagator for a degenerate light quark. We denote in bold character the spatial coordinates. With reference to Fig.~\ref{fig:diagrams} we have $x=(t_\mathrm{src},\bm{x})$, $y=(t_\mathrm{snk},\bm{x})$ and $z=(z_0,\bm{z})$. We call $t_\mathrm{sep}=|t_\mathrm{snk}-t_\mathrm{src}|$ the time separation between the two interpolating operators. The expression for the pion mass splitting eq.~\eqref{eq:midpoint} is then rewritten as
\begin{flalign}
\Delta M_\pi(\Lambda)= \frac{e^2}{2} \lim_{t_\mathrm{sep}\to \infty} \sum_{z_0=-\infty}^{+\infty} \frac{\sum_{\bm{x},\bm{y},\bm{z}} \delta^{\mu\nu}\mathcal{G}_\Lambda(z)\big[C_{\mu\nu}^\mathrm{conn}(z_0,t_\mathrm{sep},\bm{x},\bm{y},\bm{z})-C_{\mu\nu}^\mathrm{disc}(z_0,t_\mathrm{sep},\bm{x},\bm{y},\bm{z})\big]}{\sum_{\bm{x},\bm{y}}C^\mathrm{2pt}(t_\mathrm{sep},\bm{x},\bm{y})}.    
\end{flalign}
After summing over the spatial coordinates $\Delta M_\pi(\Lambda)$ is a function of only $t_\mathrm{sep}$ and the temporal separation between the two vector currents. 
\begin{table}
\begin{center}
\footnotesize{
\begin{tabular}{cccccccc}
\toprule
ID & $L^3\times T$ & $a$ [fm] &$\beta$ & $L$ [fm] & $M_\pi$ [MeV] & $M_\pi L$   \\
\midrule
\midrule
H102 & $32^3\times 96$  & 0.085 & 3.40 & 2.7 & 358 & 4.9 \\
H105 & $32^3\times 96$  & 0.085 & 3.40 & 2.7 & 280 & 3.9 \\
N101 & $48^3\times 128$ & 0.085 & 3.40 & 4.1 & 280 & 5.8 \\
C101 & $48^3\times 96$  & 0.085 & 3.40 & 	4.1 & 220 & 4.6 \\
\midrule
S400 & $32^3\times 128$ & 0.075 & 3.46 & 2.4 & 355 & 4.3 & \\
N451 & $48^3\times 128$ & 0.075 & 3.46 & 3.6 & 290 & 5.3 & \\
D450 & $64^3\times 128$ & 0.075 & 3.46 & 4.8 & 220 & 5.3 & \\
D452 & $64^3\times 128$ & 0.075 & 3.46 & 4.8 & 160 & 3.8 & \\
\midrule
N203 & $48^3\times 128$ & 0.064 & 3.55 & 3.0 & 350 & 5.4 & \\
N200 & $48^3\times 128$ & 0.064 & 3.55 & 3.0 & 290 & 4.4 & \\
D251 & $64^3\times 128$ & 0.064 & 3.55 & 4.1 & 290 & 5.9 & \\
D200 & $64^3\times 128$ & 0.064 & 3.55 & 4.1 & 200 & 4.2 & \\
\midrule
N302 & $48^3\times 128$ & 0.049 & 3.70 & 2.4 & 350 & 4.2 & \\
J303 & $64^3\times 192$ & 0.049 & 3.70 & 3.1 & 260 & 4.1 & \\
\bottomrule
\bottomrule
\end{tabular}
}
\caption{\label{tab:ensembles} CLS $n_f=2+1$ ensembles used in this work. See \cite{Bruno:2014jqa,Bali:2016umi,Djukanovic:2024cmq} and references therein for further information.}
\end{center}
\end{table}
The object appearing in the denominator is the  two-point function with the pion interpolators evaluated at fixed time slice $t_\mathrm{sep}$,
\begin{flalign}
C^\mathrm{2pt}(t_\mathrm{sep},\bm{x},\bm{y})=\Tr\big[\gamma_5 S(x,y) \gamma_5S(y,x)\big].
\end{flalign}
In the mid-point method this ensures that the limit $t_\mathrm{sep}\to\infty$ projects the pion to the ground state at zero momentum and that excited-state contamination is washed out.  At the present stage of this work we neglect the disconnected contribution since it is numerically challenging but also suppressed compared to the connected one (see for instance \cite{Feng:2021zek} where the disconnected contribution is shown to be around $1\%$ of the connected one).  Given this, we write the pion mass splitting as 
\begin{flalign}\label{eq:lattice_delta_mpi}
\Delta M_\pi(\Lambda) = \frac{e^2}{2} \lim_{t_\mathrm{sep}\to \infty} \sum_{z_0=-\infty}^{+\infty} f^\mathrm{latt}_\Lambda(z_0,t_\mathrm{sep}),
\end{flalign}
where
\begin{flalign}
f^\mathrm{latt}_\Lambda(z_0,t_\mathrm{sep}) = \frac{\sum_{\bm{x},\bm{y},\bm{z}} \delta^{\mu\nu}\mathcal{G}_\Lambda(z)\big[C_{\mu\nu}^\mathrm{conn}(z_0,t_\mathrm{sep},\bm{x},\bm{y},\bm{z})\big]}{\sum_{\bm{x},\bm{y}}C^\mathrm{2pt}(t_\mathrm{sep},\bm{x},\bm{y})}.
\end{flalign}

The relevant correlation functions have been calculated on the lattice ensembles reported in Tab.~\ref{tab:ensembles}. They were generated within the Coordinate-Lattice-Simulations (CLS) effort. 
We fix the temporal time slice of the origin to $T/2$, where $T$ is the temporal extent of the lattice, and the sink and source positions are placed symmetric w.r.t. it, that is $t_\mathrm{snk}=T/2+t_\mathrm{sep}/2$ and $t_\mathrm{snk}=T/2-t_\mathrm{sep}/2$. We have generated data for several values of $t_\mathrm{sep}$ and performed the $t_\mathrm{sep}\to \infty$ limit in a plateau sense.  As for the electromagnetic currents, we employ the local  current in the origin and the point-split definition in $z$ which, being exactly conserved on the lattice, requires no renormalization factors.  In addition, in order to account for the wrap around effects in the periodic boxed we correct the large time part of $C^\mathrm{2pt}$ by using a single-exponential fit.
\section{Long-distance contribution}
\begin{figure}
\centering
\includegraphics[width=0.48\linewidth]{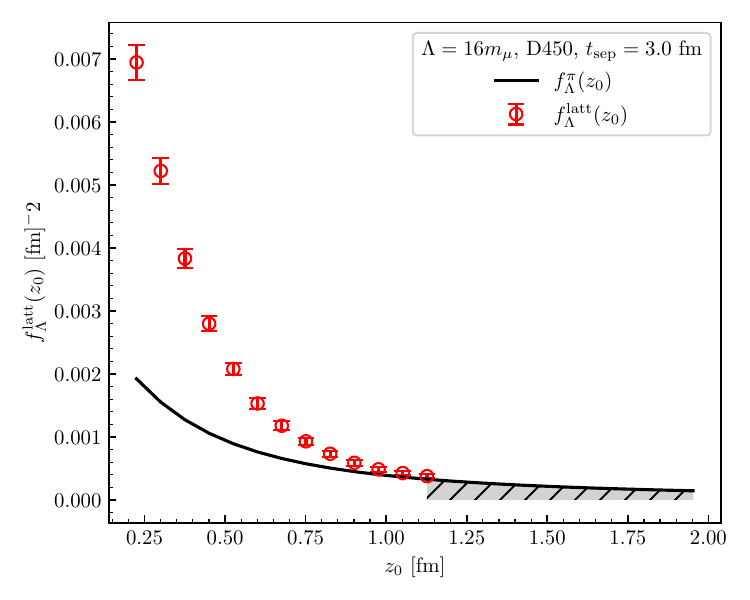}
\includegraphics[width=0.48\linewidth]{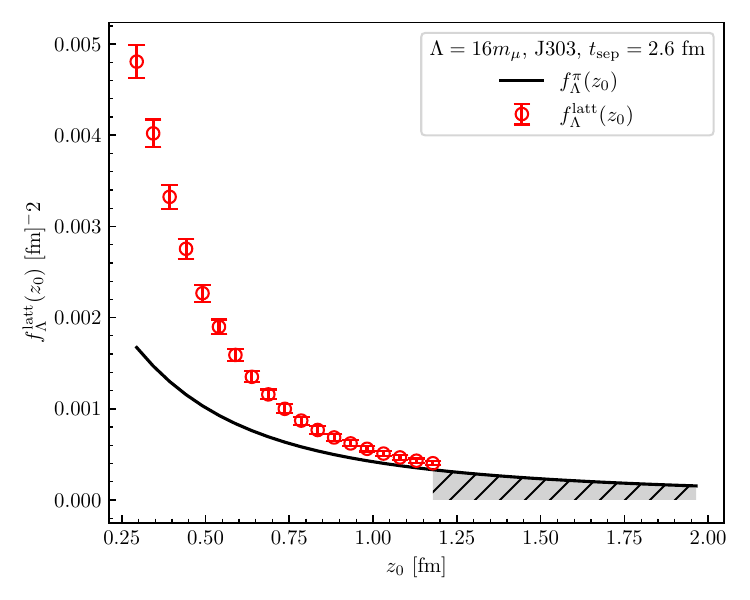}
\caption{Comparison between the lattice data and the analytical infinite volume contribution assuming only single-pion states are propagating. The plots refer to the D450 (left) and J303 (right) ensembles respectively, both at the PV-mass $\Lambda=16\,m_\mu$. The shaded area represents the region where $f_\Lambda^\pi$ replaces $f^\mathrm{latt}_\Lambda$ in evaluating the pion mass splitting.}
\label{fig:fpi}
\end{figure}
The discrete integration in eq.~\ref{eq:lattice_delta_mpi} cannot be actually extended up to $|z_0|=\infty$ since the number of available time slices is limited by $t_\mathrm{sep}$. On the other hand, most of the bulk of the finite volume effects are expected to come from the large temporal separation. To tackle the problem we, inspired by \cite{Feng:2018qpx,Feng:2021zek}, split the pion mass splitting into a short distance contribution, calculated on the lattice in the range $z_0\in[-t_c,t_c]$, and a long distance contribution ($z_0=[t_c,\infty]$) calculated analytically already in the continuum and infinite volume limit. Such analytical contribution is derived under the assumption (see \cite{Erb:2025nxk} for further details) that at large distance only single-pion states enter the intermediate states propagating between the two vector currents. The expression in infinite volume and incorporating the PV-regulated photon propagator is 
\begin{flalign}\label{eq:fpi}
f^\pi_\Lambda(z_0)= \frac{1}{4\pi^2}\int_{0}^{\infty} \dd{|\bm{p}|} |\bm{p}|^2 e^{ |z_0|(M_\pi-E_{\bm{p}})}\frac{E_{\bm{p}}+M_\pi}{E_{\bm{p}}}\times\big[F(-q)^2\big]^2\\ \nonumber
\times \bigg[\frac{e^{-|\bm{p}||z_0|}}{2|\bm{p}|} - \frac{e^{-|z_0|\sqrt{|\bm{p}|^2+\Lambda^2/2}}}{\sqrt{|\bm{p}|^2+\Lambda^2/2}} + \frac{1}{2}\frac{e^{-|z_0|\sqrt{|\bm{p}|^2+\Lambda^2}}}{\sqrt{|\bm{p}|^2+\Lambda^2}}\bigg],
\end{flalign}
where $F(-q^2)$ is the form factor (evaluated at zero spatial momentum) parametrizing the transition between pion states mediated by the electromagnetic current, 
\begin{flalign}
\bra{\pi,\bm{p}} J_\mu^\mathrm{em}(0)\ket{\pi,\bm{k}} = -i(p_\mu + k_\mu)F\big(-q^2\big),\qquad q_\mu = p_\mu-k_\mu.
\end{flalign}
In this work we employ the Vector Meson Dominance (VMD) parametrization for the form factor,
\begin{flalign}\label{eq:VMD}
F(-q^2)=\frac{1}{1+q^2/M_\mathrm{VMD}^2}.
\end{flalign}
The expression in eq.~\ref{eq:fpi} can be derived also in finite volume and it can be used to correct the lattice data for the residual finite volume effects in the range $z_0\in[-t_c,t_c]$.  Eventually the pion mass splitting is obtained as
\begin{flalign}
\Delta M_\pi (\Lambda) = \frac{e^2}{2} \bigg[ \sum_{z_0=-t_c}^{+t_c} f^\mathrm{latt}_\Lambda(z_0,t_\mathrm{sep}=\infty) +  2 \int_{t_c}^{\infty} \dd{z_0} f^\pi_\Lambda(z_0)\bigg]. 
\end{flalign}
Notice that $f_\Lambda^\pi(z_0)$ is a symmetric function w.r.t. $z_0=0$ from which we get the factor 2. This procedure is employed for all the data generated on the ensembles listed in Tab.~\ref{tab:ensembles}. The cut time $t_c$ is set around $1\sim 1.2$~fm for all the ensembles.  Fig.~\ref{fig:fpi} shows the comparison between $f^\mathrm{latt}_\Lambda(z_0)$ and $f^\pi_\Lambda(z_0)$ for the ensembles $D450$ and $J303$ with the value of the photon cutoff being $16\,m_\mu$. The good matching between the two at around $t_c=1.2~$fm is evidence for the smallness of the finite volume effects in our lattice data, a remarkable benefit of using the photon propagator in the infinite volume which automatically avoids power-law finite-size effects.

\section{Chiral-continuum extrapolation}
\begin{figure}[t]
\centering
\includegraphics[width=\linewidth]{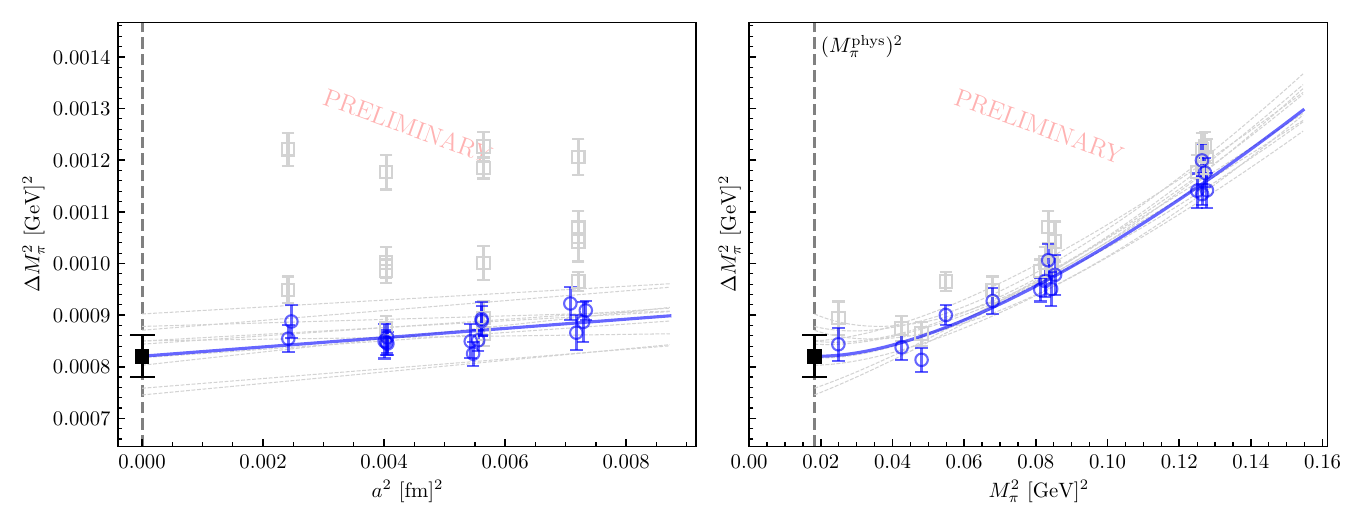}
\caption{Example of the chiral-continuum extrapolation at $\Lambda=20\,m_\mu$. The dependence on $a^2$ and on $M_\pi^2$ are shown in the left and right plot respectively. In both the plots the gray squares correspond to the original lattice data, while the blue dots correspond to the ones corrected by the fit ansatz using $M_\pi=M_\pi^\mathrm{phys}$ (left plot) and $a=0$ (right plot). In the left plot the blue points corresponding to the same lattice spacing are slightly misplaced to improve the visibility. The black square is the result of the extrapolation.}
\label{fig:chiral_cont}
\end{figure}
Since our lattice data are produced on ensembles with different lattice spacings and with pion masses heavier than the physical one, $M_\pi^\mathrm{phys}=134.98~$MeV, a chiral-continuum extrapolation to $a=0$ and $M_\pi=M_\pi^\mathrm{phys}$ is needed. For convenience, we consider the squared mass splitting 
\begin{flalign}
\Delta M_\pi^2 = M_\piplus^2-M_\pizero^2=2M_\pi\Delta M_\pi + \order{\alpha^2_\mathrm{em}}.
\end{flalign}
Inspire by ChPT  (see for instance \cite{Hayakawa:2008an,Gasser:2003hk}) and by noticing that the pion mass splitting at $\order{\alpha_\mathrm{em}}$ is purely due to electromagnetism, namely there is no dependence on the difference of light quark masses, we  come up with the following fit ansatz to model the dependence on the lattice spacing and on the pion mass,
\begin{flalign}
\Delta M_\pi^2(\Lambda,a,M_\pi)=e^2 \bigg[c_0 +c_a a^2  + c_\pi^{(1)} M_\pi^2 + c_\pi^{(2)} M_\pi^2 \log\bigg(\frac{M_\pi^2}{\mu^2}\bigg)\bigg].
\end{flalign}
Here we set $\mu=4\pi f_\pi\sim 1.2$~GeV. This fit ansatz provides a good description of the dependence upon lattice artifacts and pion mass with the current statistical precision of our data  points as shown in Fig.~\ref{fig:chiral_cont} for $\Lambda=20\,m_\mu$. After performing the chiral-continuum extrapolation individually for all the values of the photon mass we come back to pion mass splitting by using the relation $\Delta M_\pi = \Delta M_\pi^2/2M_\pi^\mathrm{phys}+\order{\alpha_\mathrm{em}^2}$.

\section{Comparison with phenomenology and $\Lambda\to \infty$ limit}
\begin{figure}
\centering
\includegraphics[width=\linewidth]{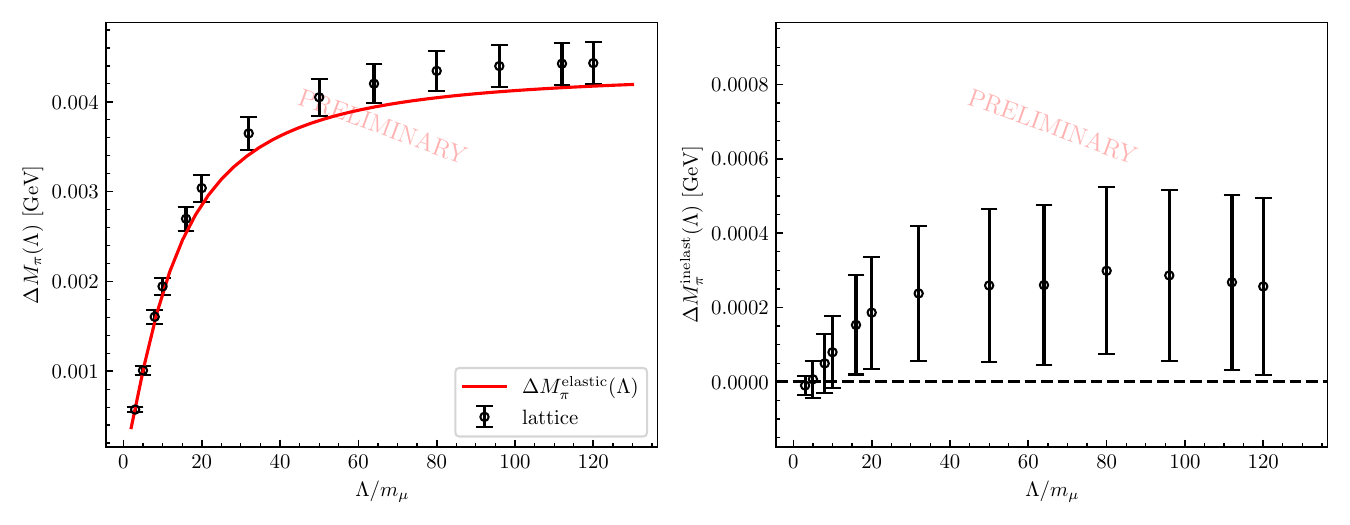}
\caption{\emph{Left}: our lattice determination for $\Delta M_\pi(\Lambda)$ (black dots) after chiral-continuum extrapolation  for different values of the photon mass compared to the elastic contribution (solid red line) determined from the forward Compton amplitude. \emph{Right}: inelastic contribution and $\Lambda$ dependence. }
\label{fig:phenomenology}
\end{figure}

An interesting phenomenological aspect is the separation of the elastic contribution (in which only pions appear in the intermediate states) from the inelastic one which, instead, contains the contribution of all the other states. Besides the phenomenological implications, this separations is particular convenient to address the $\Lambda \to \infty$ limit. The pion mass splitting plays a special role in this regard being UV-finite, but for a general hadron the mass splitting is divergent in the limit $\Lambda\to \infty$, when the proper counterterms are not taken into account. This divergence is expected to be given by the inelastic part and it is therefore particularly important to separate the elastic contribution from it. Of course, the lattice data contains both contributions. The elastic contribution can be defined dispersively by relying on the Cottingham formula which connects the electromagnetic self energy to the forward Compton tensor. We follow \cite{Stamen:2022uqh}  and write the elastic electromagnetic correction to the pion mass as
\begin{flalign}\label{eq:compton}
(\Delta M_\pi^\mathrm{elast})^2 = \frac{ie^2}{2}\int \frac{\dd[4]{k}}{(2\pi)^4}\frac{T^{\mu,\mathrm{elast}}_\mu}{k^2+i\varepsilon},
\end{flalign}
where $T^{\mu}_\mu$ is the elastic forward Compton tensor which is expressed in terms of the meson form factor. For the pion mass spltting we make use of the VMD parametrization of eq.~\ref{eq:VMD} with $M_\mathrm{VMD}=M_\rho=0.77~$GeV. We modified eq.~\ref{eq:compton} to include the regulated propagator of eq.~\ref{eq:PV_propagator} and obtain $\Delta M_\pi^\mathrm{elast}(\Lambda)$ for the values of the photon mass used in this work and in particular $\Delta M_\pi^\mathrm{elast}(\infty)=4.33~$MeV. The chirally-continuum extrapolated lattice data are compared to the elastic constribution in the left plot of Fig.~\ref{fig:phenomenology}. The inelastic part can be obtained by subtracting $\Delta M_\pi^\mathrm{elast}(\Lambda)$ to the lattice data at fixed $\Lambda$,
\begin{flalign}
\Delta M_\pi^\mathrm{inelast}(\Lambda)=\Delta M_\pi^\mathrm{latt}(\Lambda)-\Delta M_\pi^\mathrm{elastic}(\Lambda).
\end{flalign}
The result for $\Delta M_\pi^\mathrm{inelast}(\Lambda)$ is shown in the right plot of Fig.\ref{fig:phenomenology}. As expected in this case, the inelastic contribution shows no divergence and the on-set of the plateau is reached quite fast, at around $\Lambda=20\,m_\mu$, in correspondence of which our lattice calculation provides $\Delta M_\pi^\mathrm{inelast}(\Lambda=\infty)\simeq\Delta M_\pi^\mathrm{inelast}(\Lambda=20\,m_\mu)=0.19(15)~$MeV. Adding back the elastic contribution in the $\Lambda \to \infty$ limit gives $\Delta M_\pi^\mathrm{latt}(\Lambda=\infty)=4.52(15)~$MeV, which is in very good agreement with the experimental measurement $\Delta M_\pi^\mathrm{exp}=4.5936(5)~$MeV \cite{ParticleDataGroup:2024cfk}.
\section{Conclusions}
In this work we have determined the pion mass splitting in lattice QCD using a photon propagator already in infinite volume and with an additional regulator given by the Pauli-Villars mass $\Lambda$. Combining the infinite-volume photon propagator with the description of the long-distance contribution already in infinite-volume avoids power-law finite-volume effects. The introduction of the scale $\Lambda$ facilitates the separation between the elastic and the inelastic part. The former is determined phenomenologically by using a Cottingham formula with the modified PV-regulated photon propagator. The latter shows little dependence on $\Lambda$ allowing for an easy $\Lambda\to \infty$ extrapolation. This operation gives the preliminary result $\Delta M_\pi^\mathrm{latt}=4.52(15)~$MeV, already extrapolated to the continuum and to the physical point, which is well in agreement with the experimental one. Even though the introduction of the additional scale $\Lambda$ is not strictly necessary for the case of the pion, this investigation validates the method (first proposed in \cite{Biloshytskyi:2022ets}) and provides useful insight for future applications to other observables, such as the proton/neutron mass splitting and the anomalous magnetic moment of the muon, where the application of this formalism has already been started \cite{Erb:2025nxk}.
\section{Acknowledgements}
 Calculations for this project were performed on the HPC cluster “Mogon II” at JGU Mainz. We are grateful to our colleagues in the CLS initiative for sharing ensembles.
We acknowledge the support of Deutsche Forschungsgemeinschaft (DFG, German Research Foundation) through project JRP (No.~458854507) of the research unit FOR 5327 “Photon-photon interactions in the Standard Model and beyond exploiting the discovery potential from MESA to the LHC”, and through the Cluster of Excellence “Precision Physics, Fundamental Interactions and Structure of Matter” (PRISMA+ EXC 2118/1) funded within the German Excellence Strategy (project ID 39083149). 
\bibliographystyle{JHEP}
\bibliography{bibliography}

\end{document}